# Physics-Constrained Diffusion Reconstruction with Posterior Correction for Quantitative and Fast PET Imaging

Yucun Hou, Fenglin Zhan, Chenxi Li, Ziquan Yuan, Haoyu Lu, Yue Chen, Yihao Chen, Kexin Wang, Runze Liao, Haoqi Wen, Ganxi Du, Jiaru Ni, Taoran Chen, Jinyue Zhang, Jigang Yang, Jianyong Jiang

*Abstract*—**Deep learning–based reconstruction of positron emission tomography (PET) data has gained increasing attention in recent years. While these methods achieve fast reconstruction, concerns remain regarding quantitative accuracy and the presence of artifacts, stemming from limited model interpretability, data-driven dependence, and overfitting risks. These challenges have hindered clinical adoption. To address them, we propose a conditional diffusion model with posterior physical correction (PET-DPC) for PET image reconstruction. An innovative normalization procedure generates the input Geometric TOF Probabilistic Image (GTP-image), while physical information is incorporated during the diffusion sampling process to perform posterior scatter, attenuation, and random corrections. The model was trained and validated on 300 brain and 50 whole-body PET datasets, a physical phantom scanned using a Siemens Biograph Vision PET/CT at The First Affiliated Hospital of the University of Science and Technology of China, and 20 simulated brain datasets. PET-DPC produced reconstructions closely aligned with fully corrected OSEM images, outperforming end-to-end deep learning models in quantitative metrics and, in some cases, surpassing traditional iterative methods. The model also generalized well to out-of-distribution (OOD) data. Compared to iterative methods, PET-DPC reduced reconstruction time by ~50% for brain scans and ~85% for whole-body scans. Ablation studies confirmed the critical role of posterior correction in implementing scatter and attenuation corrections, enhancing reconstruction accuracy. Experiments with physical phantoms further demonstrated PET-DPC's ability to preserve background uniformity and accurately reproduce tumor-to-background intensity ratios. Overall, these results highlight PET-DPC as a promising approach for rapid, quantitatively accurate PET reconstruction, with strong potential to improve clinical imaging workflows.**

## I. INTRODUCTION

Positron emission tomography (PET) is a non-invasive functional medical imaging technology widely utilized in clinical diagnosis[1-6]. The introduction of iterative reconstruction algorithms, such as maximum likelihood expectation maximization (MLEM) and its accelerated variant ordered subset expectation maximization (OSEM), has substantially improved PET image quality [7, 8]. However, despite these advances, iterative methods remain computationally intensive and time-consuming, particularly for whole-body scans that incorporate Time-of-Flight (TOF) and Depth-of-Interaction (DOI) information, as well as Continuous Bed Motion (CBM) acquisition. These limitations pose significant challenges to clinical workflow

Recent advancements in deep learning have led to widespread application in PET image processing, including image denoising, scatter correction, attenuation correction, multimodal translation, and computer-aided diagnosis, among others[9-14]. By learning optimal parameters to solve ill-posed inverse problems and enabling direct, one-step image reconstruction, deep learning offers a clear advantage over conventional iterative methods. Zhu et al. reconceptualized the reconstruction process as a mapping between the detected data domain and the image domain, and introduced the AUTOMAP framework[15], which reconstructs PET image from the attenuation-corrected two-dimensional sinogram data, demonstrating the feasibility of deep learning–based image reconstruction. Häggström et al. proposed DeepPET[16], an end-to-end PET image reconstruction approach employing a VGG16-based encoder–decoder architecture trained on simulated sinogram data. Beyond sinogram-based strategies, other approaches leverage simple back-projection of raw data to generate network inputs. For instance, Whiteley et al. utilized TOF information to project lines of response (LORs) into a Most Likely Annihilation Position histogrammer (histo-image).

This work was supported in part by the National Natural Science Foundation of China (No. 12475336), Beijing Nova Program (No. 20230484413). (Yucun Hou, Fenglin Zhan and Chenxi Li are co-first authors. Corresponding author: Jianyong Jiang, Jigang Yang.)

Yucun Hou, Chenxi Li, Ziquan Yuan, Yue Chen, Yihao Chen, Kexin Wang, Runze Liao, Haoqi Wen, Ganxi Du, Jiaru Ni, Jianyong Jiang are with School of Physics and Astronomy, Beijing Normal University, Beijing 100875. China and with Key Laboratory of Beam Technology of Ministry of Education, Beijing Normal University, Beijing 100875. China (e-mail: hou9021@mail.bnu.edu.cn; jianyong@bnu.edu.cn). Taoran Chen and Jinyue Zhang are with School of Mathematical Sciences, Beijing Normal University, Beijing 100875. China.(email:) Jigang Yang is with Department of Nuclear Medicine, Beijing Friendship Hospital, Capital Medical University, 95 Yong An Road, Xi Cheng District, Beijing 100050, China.(email: ). Fenglin Zhan is with Department of Nuclear Medicine，The First Affiliated Hospital of USTC, Division of Life Sciences and Medicine, University of Science and Technology of China, Hefei, Anhui, 230001, China：Wuxi School of Medicine, Jiangnan University, No. 1800, Lihu Avenue, Wuxi, 214000, China (e-mail: ). Haoyu Lu is with Department of Biomedical Engineering, University of Melbourne, Melbourne VIC 3010, Australia (e-mail: ).

This histo-image, combined with the corresponding attenuation map, was used as input to a 3D U-Net model, resulting in the efficient PET reconstruction framework FastPET[17]. Yang et al. further advanced this direction by introducing a multitask learning (MTL) strategy[18], in which roughly corrected sinograms generate initial back-projected images as inputs. A noisy PET image serves as a weakly supervised conditional training target, while an auxiliary task predicts anatomical images to suppress noise propagation.

Due to the data-driven nature of deep learning models, supervised approaches, including those with weak supervision, are highly sensitive to the distribution of the training data. This sensitivity often leads to limited generalizability when models are applied to data outside the original training domain[19]. Specifically, variations in imaging characteristics or anatomical regions can introduce domain shifts, leading to degraded performance on out-of-distribution (OOD) cases. For instance, a model trained exclusively on brain PET datasets may fail to accurately reconstruct whole-body or physical phantom data, as demonstrated in the experimental section of this study. Moreover, end-to-end approaches generally do not explicitly model the physical processes inherent in PET imaging. In conventional reconstruction, numerous factors must be accounted for, including spatially varying system resolution, attenuation correction, scatter correction, random correction, and normalization. In contrast, deep learning–based reconstruction methods implicitly rely on the network to learn and compensate for all these effects. Given the black-box nature of neural networks, it remains unclear whether such essential corrections are consistently and accurately performed. These limitations collectively raise concerns regarding the quantitative reliability of PET images reconstructed solely through data-driven deep learning approaches.

To reduce dependence on training datasets and inspired by the deep image prior (DIP) framework[20], Gong et al. incorporated a deep learning model into the reconstruction process in a training data-free mode[19]. Their DIPRecon model reformulates the log-likelihood objective as a neural network parameter optimization problem and solves it using the Alternating Direction Method of Multipliers (ADMM) algorithm. However, the additional inverse process incurs computational costs, and the hyperparameter $\rho$ of the ADMM algorithm is typically difficult to tune.

Hashimoto et al. further extended the DIP concept by optimizing network parameters through a loss function defined between the network output and measured data in the sinogram domain[21]. Compared to the nested two-level iterative scheme in DIPRecon, their approach reduces one level of iteration. Nonetheless, this approach simply models PET forward projection as the product of the activity distribution and a projection matrix. In PET imaging, the probability that gamma rays emitted from radiotracer decay are detected as coincidence events follows a Poisson distribution. This physical process is further affected by scatter and random coincidences. They preemptively removed scatter events and did not consider random coincidences. While scatter data could be removed in simulated data, the pre-estimation of scatter for real data was inaccurate. Therefore, this approach represented an incomplete physical model for PET reconstruction. In parallel, Siqi et al. proposed two deep learning-based kernel methods[22, 23], which achieved superior dynamic PET reconstruction compared with other kernel methods and DIP-based approaches. However, these methods incur higher computational costs, and their reliance on prior images limits their applicability to static PET imaging.

More recently, diffusion models[24, 25] have demonstrated remarkable success in PET image processing[26, 27]. For inverse problems, Chung et al. proposed the Diffusion DPS method, which integrates measurement data into the reverse diffusion step to constrain the sampling process[28]. Building upon these advancements, we propose a diffusion-based PET reconstruction framework with posterior correction (PET-DPC). In this approach, list-mode raw data are first back-projected to generate a Geometric TOF Probabilistic Image (GTP-image), providing a coarse estimate of the activity distribution that integrates both TOF and geometric information. A conditional DDPM is then trained with the GTP-image as auxiliary input. During the reverse diffusion process, measurement data are incorporated as posterior information to perform physical corrections.

The experimental dataset utilized in this study comprises 300 brain scans, 50 whole-body scans, 20 simulated brain scans, and one physical phantom. Notably, the simulated data and the physical phantom were excluded from training and instead used as OOD cases to assess model generalization, reconstruction uniformity, and tumor uptake accuracy.

## II. Material And Method

### A. Conditional DDPM

Within the DDPM framework, the forward diffusion process is modeled as a fixed Markov chain that gradually transforms the data distribution $p(x_0)$ into a Gaussian distribution $p(x_T)$ by sequentially adding Gaussian noise, as shown by:

$$p(x_{1:T} \mid x_0) := \prod_{t=1}^{T} p(x_t \mid x_{t-1}),$$

$$p(x_t \mid x_{t-1}) := \mathcal{N}(x_t; \sqrt{\alpha_t} x_{t-1}, (1 - \alpha_t)I) \quad (1)$$

The noise schedule satisfies $0 < \alpha_T < \alpha_{T-1}, \dots, \alpha_1 < 1$. By reparameterization, the distribution of $x_t$ at any time $t$ in the forward process, conditioned on $x_0$, can be computed as:

$$p(x_t \mid x_0) := \mathcal{N}(x_t; \sqrt{\bar{\alpha}_t} x_0, (1 - \bar{\alpha}_t)I)$$

$$x_t = \sqrt{\bar{\alpha}_t} x_0 + \sqrt{1 - \bar{\alpha}_t} \epsilon(x_t, t), \qquad \epsilon(x_t, t) \sim \mathcal{N}(0, I) \quad (2)$$

Here, $\bar{\alpha}_t := \prod_{s=1}^{t} \alpha_s$ denotes the cumulative product of noise scales up to timestep $t$. For reverse process, the posterior distribution $p(x_{t-1} \mid x_t)$ is derived using Bayes' theorem as follows:

$$p(x_{t-1} \mid x_t, x_0) = \frac{p(x_t \mid x_{t-1}, x_0) p(x_{t-1} \mid x_0)}{p(x_t \mid x_0)}$$

$$= \mathcal{N}(x_{t-1}; \mu(x_t, x_0), \sigma_t^2 I) \quad (3)$$

The mean $\mu(x_t, x_0)$ and the variance $\sigma_t{}^2$ are explicitly defined as follows:

$$\mu(x_t, x_0) = \frac{\sqrt{\alpha_t}(1 - \bar{\alpha}_{t-1})}{1 - \bar{\alpha}_t}x_t + \frac{\sqrt{\bar{\alpha}_{t-1}}(1 - \alpha_t)}{1 - \bar{\alpha}_t}x_0$$

$$\sigma_t{}^2 = \frac{1 - \bar{\alpha}_{t-1}}{1 - \bar{\alpha}_t}(1 - \alpha_t) \quad (4)$$

Then, a neural network is trained to predict the distribution $p_\theta(x_{t-1} \mid x_t)$, aiming to approximate the true posterior $p(x_{t-1} \mid x_t)$. By substituting the reparameterized form of (2), $x_0$ is given by $(x_t - \sqrt{1 - \bar{\alpha}_t}\epsilon(x_t, t))/\sqrt{\bar{\alpha}_t}$. The mean of the posterior distribution $p(x_{t-1} \mid x_t)$ is given by:

$$\mu(x_t, t) = \frac{1}{\sqrt{\alpha_t}}x_t - \frac{1 - \alpha_t}{\sqrt{\alpha_t}\sqrt{1 - \bar{\alpha}_t}}\epsilon(x_t, t) \quad (5)$$

Given $x_t$ and $t$, the only unknown variable is the noise $\epsilon(x_t, t)$. Accordingly, the training objective is to minimize the discrepancy between the predicted noise $\epsilon_\theta(x_t, t)$ produced by the neural network and the true noise $\epsilon(x_t, t)$. In Conditional DDPM, additional prior information $x_C$ may be incorporated during the noise prediction process. In this case, the predicted noise becomes $\epsilon_\theta(x_t, t, x_C)$. This objective is typically formulated as the mean squared error (MSE) between the predicted and true noise over the training dataset:

$$\mathcal{L}_\theta = \mathbb{E}_{x_{0,\epsilon}}\big[\|\epsilon_\theta(x_t, t, x_C) - \epsilon(x_t, t)\|^2\big] \quad (6)$$

Based on (3), sampling can be performed from the posterior distribution $p_\theta(x_{t-1} \mid x_t)$, gradually sampling from $x_T$ to $x_0$ through the following iterative process:

$$x_{t-1} = \mu_\theta(x_t, t, x_C) + \sigma_t z, \quad \text{where } z \sim \mathcal{N}(0, \mathrm{I}) \quad (7)$$

### B. SDE Formulation of DDPM and Posterior Correction

As $T \to \infty$ and $t$ becomes continuous, the discrete DDPM forward process in (2) converges to the following form of a stochastic differential equation (SDE):

$$dx = -\frac{1}{2}\beta(t)x_t dt + \sqrt{\beta(t)}d\bar{w} \quad (8)$$

This represents a form of VP-SDE, with the corresponding reverse process given by:

$$dx = \left[-\frac{1}{2}\beta(t)x_t - \beta(t)\nabla_{x_t}\log p_t(x_t)\right]dt + \sqrt{\beta(t)}d\bar{w} \quad (9)$$

Here, $\nabla_{x_t}\log p_t(x_t)$ corresponds to the score function in conditional DDPM.

In PET imaging, the detector counts follow a Poisson distribution, and the mean of the detected coincidence data $\bar{y}$ can be described as:

$$\bar{y}_i = \sum_j P_{ij}x_j + \bar{r}_i + \bar{s}_i \quad (10)$$

Here, $i$ denotes the i-th detected coincidence event, and $j$ refers to the j-th voxel in the reconstruction region. $\bar{r}_i$ and $\bar{s}_i$ represent the means of random and scatter events, respectively. As a result of the presence of scatter and random events, PET reconstruction becomes an ill-posed inverse problem. The probability of detecting each pair of coincident events can be written as:

$$p(y \mid x_0) = \prod_i \frac{\exp(-\bar{y}_i) * \bar{y}_i{}^{y_i}}{y_i!} \quad (11)$$

Where $x_0$ represents the set of detected pixels $x_j$ within the reconstruction region. If the detected event $y$ is used as posterior information, the score function in (9) can be represented as the posterior probability:

$$dx = \left[-\frac{1}{2}\beta(t)x_t - \beta(t)\nabla_{x_t}\log p_t(x_t \mid y)\right]dt + \sqrt{\beta(t)}d\bar{w} \quad (12)$$

Based on Bayes' theorem, the following can be derived:

$$\nabla_{x_t}\log p_t(x_t \mid y) = \nabla_{x_t}\log p_t(x_t) + \nabla_{x_t}\log p_t(y \mid x_t) \quad (13)$$

The first term corresponds to the score function. For the second term, based on the conclusion derived in [28], it follows that $\nabla_{x_t}\log p_t(y \mid x_t) \simeq \nabla_{x_t}\log p_t(y \mid \hat{x}_0)$. Based on (11), the logarithmic form is given by:

$$\log p(y \mid \hat{x}_0) = \sum_{i=1}^{M} y_{ik\tau}\log \bar{y}_{ik\tau} - \bar{y}_i - \log y_{ik\tau}! \quad (14)$$

This term allows for correction of the sampling process, serving as the correction term. Based on the ancestral sampling method in [24], the solution can be derived. However, unlike DPS, the sampling process is not directly adopted as ancestral sampling. Instead, equation (14) is used to compute the gradient with respect to $\hat{x}_0$, and measurement data is employed to update $\nabla_{x_0}\log p_t(y \mid \hat{x}_0)$. A Predictor-Corrector (PC) sampling strategy, as introduced in [25], is then used for sampling. The updated $\hat{x}_0$ is used as the Corrector, followed by the use of ancestral sampling as the Predictor, as outlined in Algorithm 1.

---

**Algorithm 1** Diffusion Posterior Correction

**Require:** $N$, $y$
1: $x_{T-1} \sim \mathcal{N}(\mathbf{0}, \mathbf{I})$
2: **for** $i = T - 1$ **to** $0$ **do**
3: $\quad \hat{s} \leftarrow s_\theta(x_i, i)$
4: $\quad \hat{x}_0' \leftarrow \frac{1}{\sqrt{\bar{\alpha}_i}}(x_i + (1 - \bar{\alpha}_i)\hat{s})$
5: $\quad \hat{x}_0 \leftarrow \frac{c_{IM}\hat{x}_0'}{\Sigma_l P_l}\sum_l \frac{y_l}{\bar{y}_l}P_l$
6: $\quad z \sim \mathcal{N}(\mathbf{0}, \mathbf{I})$
7: $\quad x_{i-1} \leftarrow \frac{\sqrt{\alpha_i}(1 - \bar{\alpha}_{i-1})}{1 - \bar{\alpha}_i}x_i + \frac{\sqrt{\bar{\alpha}_{i-1}}(1 - \alpha_i)}{1 - \bar{\alpha}_i}\hat{x}_0 + \frac{1 - \bar{\alpha}_{i-1}}{1 - \bar{\alpha}_i}(1 - \alpha_i)z$
8: **end for**
9: **return** $x_0$

---

Step 4 corresponds to Tweedie's approach. In Step 5, $i$ refers to the $l$ in (10), distinguished from the sampling step index. Since the prediction of the score function incorporates the additional $x_C$, the number of sampling steps is set to 5, with 5 posterior corrections also performed. As a result of the alternation between the correction and predictor processes, the correction Step 5 primarily applies the attenuation and scatter corrections that were not applied during the predictor process. This is further validated by the subsequent experimental results.

### C. Geometric TOF Probabilistic Image Reconstruction

The Geometric TOF Probabilistic Image (GTP-image) is obtained through a comprehensive probabilistic back-projection process that incorporates both TOF and geometric information. For each list-mode line of response (LOR), the annihilation position is estimated in the image space using TOF information. A Gaussian weighting function, determined by the system's timing resolution, is applied along the LOR to model the spatial uncertainty of the annihilation event. In parallel, a geometric weighting function is introduced to quantify the overlap between the LOR and each voxel, thereby accounting for the intersection volume. The contribution from each LOR to the image is thus jointly weighted by the TOF Gaussian and the geometric term, and further scaled by a precomputed normalization factor that accounts for detector efficiency. To eliminate the effect of positional sensitivity, the final image is normalized by a sensitivity map. The GTP-image is calculated as:

$$I_v = 1 \times N_L \times w_{\text{TOF}}(v, L) \times w_{\text{geo}}(v, L)$$

$$w_{\text{TOF}}(v, L) = \exp\left(-\frac{d_{\parallel}^2}{2\sigma_t^2}\right)$$

$$w_{\text{geo}}(v, L) = V(d_{\perp}, R, r) \qquad (15)$$

Where $N_L$ is the normalization factor that primarily accounts for detector efficiency; $w_{\text{TOF}}$ represents the TOF probability for voxel $v$, $d_{\parallel}$ denotes the projection distance of voxel $v$ along the LOR $L$ direction, $\sigma_t$ is computed based on the system's time resolution; $w_{\text{geo}}$ reflects the geometric overlap between the voxel and the LOR. Finally, the resulting image is normalized by the sensitivity image to obtain the GTP image. Fig. 1 presents GTP-images of the NEMA phantom, representative brain cases and one whole-body case.

$$GTP_{\text{map}}(v) = \sum_L I_v / S_v \qquad (16)$$

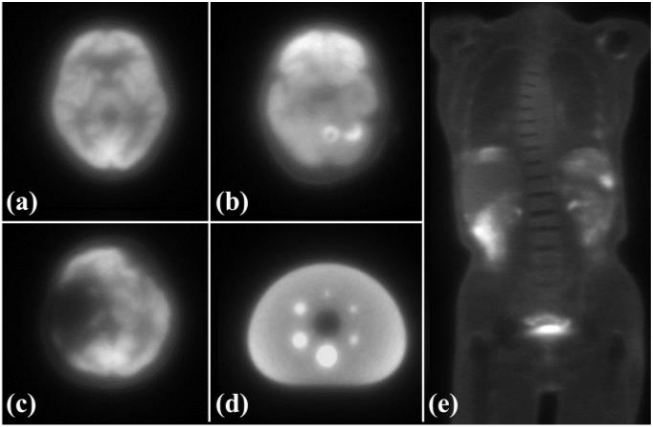

**Fig. 1.** Five representative GTP-images: (a) normal brain case; (b) tumor brain case; (c) brain case with patchy uptake defect; (d) NEMA phantom; and (e) whole-body case.

### D. Intensity Matching

Given the addition of Gaussian noise with zero mean and unit variance in the diffusion model, the output PET image is constrained to a normalized intensity range of −1 to 1. However, during posterior correction with list-mode data, the process relies on the absolute image intensity, as it directly influences the estimation of scattered events. To address this issue, an intensity-matching strategy is employed. Specifically, the pixel values in the normalized image are mapped to approximate physical activity levels based on the relationship between the injected dose and the total reconstructed activity. This is accomplished by multiplying the $\hat{x}_0'$ obtained in step 4 of

Algorithm 1 by a pre−calculated coefficient, $C_{IM}$. The approximate activity image, $C_{IM}\hat{x}_0'$, is then used for subsequent intensity-based physical correction.

## III. EXPERIMENTS

### A. Dataset and Implementation

To evaluate the performance of the proposed PET-DPC method, 300 brain scan datasets and 50 whole-body scan datasets were collected using a Siemens Biograph Vision PET/CT scanner at the First Affiliated Hospital of the University of Science and Technology of China between November 2023 and December 2024. For additional quantitative assessment, experiments were also performed on a physical phantom scanned on the same system, along with 20 simulated brain datasets generated using the same scanner model.

All data were reconstructed using the GPU-accelerated Bayesian penalized likelihood algorithm, QuanTOF[29], which incorporates time-of-flight (TOF) information and comprehensive correction techniques. QuanTOF-reconstructed images served as training labels during the training phase of the conditional diffusion model and as the ground truth for evaluating reconstruction performance.

*1) Clinical brain scan datasets:* The brain datasets were obtained from patients with an average age of 58 years and an average weight of 63.3 kg. All patients underwent standard clinical whole-body PET scans, followed by dedicated head scans after injection of 282.5 ± 52.1 MBq of ¹⁸F-fluorodeoxyglucose (¹⁸F-FDG). Each head scan lasted for 120 seconds. The head datasets were divided into 200 cases for training and 100 cases for validation. Among these, 12 cases included brain tumors, which were particularly valuable for assessing the model's ability to preserve the structural integrity of clinically relevant regions and to maintain contrast-to-noise ratio (CNR). These complex brain images served as the primary dataset for rigorous evaluation of PET-DPC's capability to reconstruct fine structural details with high fidelity.

*2) Clinical whole-body scan datasets:* The whole-body datasets were acquired from patients with an average age of 57 years and an average weight of 64.2 kg. All patients were administered ¹⁸F-FDG as the radiotracer, with a dose of 279.3 ± 45.4 MBq. Scans were performed in continuous bed motion (CBM) mode at a speed of 2.2 mm/s. The relatively high bed speed resulted in shorter acquisition times per bed position, introducing substantial noise in the reconstructed images. This acquisition setting was specifically chosen to evaluate the robustness of the proposed model under high-noise conditions. The whole-body datasets were partitioned into 45 cases for training and 5 for validation. Moreover, the varying scan geometry introduced by CBM scanning posed additional challenges for list-mode-based reconstruction. Specifically, the QuanTOF framework required several hours to reconstruct a single whole-body image due to the increased complexity of attenuation and scatter correction in CBM-driven list-mode data.

*3) Physical phantom dataset:* The physical phantom experiment was conducted using the NEMA International

Electrotechnical Commission (IEC) body phantom on a clinical Siemens Biograph Vision PET/CT scanner. The phantom was filled with 1.41 mCi of $^{18}$F-FDG and contained six tumor spheres with diameters of 10, 13, 17, 22, 28, and 37 mm, arranged to achieve a tumor-to-background activity ratio of 4:1. A 5-minute scan was performed, during which 139,229,872 prompt coincidence events were recorded.

*4) Simulation datasets:* The 20 simulated brain datasets were generated from the BrainWeb database to create PET phantom images[30, 31]. The anatomical model consists of 12 distinct regions, including gray matter, white matter, and cerebrospinal fluid. Activity values were assigned to these tissues based on the approach described in [32]. The Siemens Biograph Vision PET/CT scanner geometry was modeled in GATE[33, 34] to simulate a 120-second acquisition, yielding an average of 102,682,837 prompt coincidence events.

### B. Implementation Details

The experiments were conducted on an Ubuntu 22.04 LTS system with an Intel Xeon Gold 6226R CPU, 256GB of RAM, and three NVIDIA GeForce RTX 4090 GPUs, each with 24GB of memory. Python was used for program development, with the PyTorch deep learning framework employed for model implementation. For the brain datasets, each image had a volume of $256 \times 256 \times 256$ voxels. For the whole-body datasets, the in-plane size was $384 \times 384$ voxels, with the axial length ranging from 790 to 848 slices depending on the patient's height. All image data underwent min–max normalization to scale the voxel values to the range of $[-1, 1]$, ensuring compatibility with the input requirements of the diffusion model.

During the supervised training phase, a 2.5D strategy was employed to improve the continuity and completeness of axial information. The model was trained with a batch size of 6 using the Adam optimizer, with an initial learning rate of $1 \times 10^{-5}$ and a final learning rate of $1 \times 10^{-8}$, following a polynomial decay schedule. All timing comparisons were conducted using three GPUs in parallel.

### C. Evaluation Metrics

For the clinical data, three conventional metrics—Peak Signal-to-Noise Ratio (PSNR), Structural Similarity Index (SSIM), and Normalized Root Mean Square Error (NRMSE)—were employed to evaluate the quality of the reconstructed images. To quantitatively assess the model's ability to accurately reconstruct tumor uptake, the Contrast-to-Noise Ratio (CNR)[11, 35], which measures the relative contrast between the tumor region and the surrounding background, was also included. Tumor and background regions were selected using a threshold segmentation method, encompassing the entire tumor in the axial direction. The CNR was calculated as follows:

$$CNR = \frac{|\mu_t - \mu_b|}{\sigma_b} \qquad (17)$$

Here, $\mu_t$ and $\mu_b$ represent the mean intensities of the tumor and background regions, respectively, while $\sigma_b$ denotes the standard deviation of the background intensity.

To evaluate the uniformity of the reconstructed images and the recovery of uptake ratios, the contrast recovery coefficient (CRC) and background variability (BV) were calculated in the physical phantom study with the NEMA NU 2-2007 standard. The calculation methods are as follows:

$$CRC = \frac{\frac{c_d}{c_{B,d}} - 1}{\frac{a}{a_B} - 1} \times 100(\%) \qquad (18)$$

Where $c_d$ represents the average emission counts within the tumor region of interest (ROI) of diameter $d$, and $c_{B,d}$ denotes the average emission counts within the background ROI of diameter $d$. Here, $a$ is the activity concentration in the tumor spheres, while $a_B$ corresponds to the activity concentration in the background. The tumor-to-background ratio, defined as $a/a_B$, is equal to 4:1. The background variability (BV) quantifies the image noise within circular ROIs of diameter $d$ and is defined as follows:

$$BV = \frac{s_d}{c_{d}} \times 100(\%) \qquad (19)$$

Herein, $s_d$ refers to the standard deviation of emission counts in the background ROIs with diameter $d$.

For the simulated data, the activity ratio between gray matter and white matter was set to 3.846:1. In the activity maps, masks for the gray matter and white matter regions were selected. These masks were then applied to the reconstructed images to calculate the average intensity ratio in each region, which was compared to the predefined value. This ratio serves as a measure of contrast recovery performance. Additionally, the coefficient of variation (CV) was computed separately for the gray and white matter regions to evaluate noise suppression. The CV was calculated using the standard deviation $\sigma_{VOI}$ and the mean $\mu_{VOI}$ of the region of interest (ROI), and is defined as follows:

$$CV = \frac{\sigma_{ROI}}{\mu_{ROI}} \times 100\% \qquad (20)$$

### D. Compared Methods

The proposed PET-DPC method was compared with two baseline models. The first was the histo-image based supervised learning model, FastPET, in which a 3D U-Net was trained using both the histo-image and the attenuation map as dual-channel inputs, following the procedure described in the original publication. The second comparison model was a Conditional-DDPM without posterior correction. Aside from the omission of the correction step during sampling, its configuration is identical to that of PET-DPC.

All methods were trained using QuanTOF-reconstructed images as the training labels. For real brain and simulated brain datasets, QuanTOF was configured with 3 iterations and 5 subsets, while for the CBM whole-body datasets, 30 iterations and 1 subset were used.

## IV. RESULTS

### A. Clinical Brain Study

Fig. 2 presents a visual comparison of 12 representative tumor-containing cases selected from the 100 validation datasets, along with residual maps between each method and the QuanTOF reconstruction. Visually, the images reconstructed by FastPET appear smoother and lack structural detail, with some tumors exhibiting shape and edge discrepancies compared to QuanTOF, resulting in notable residuals. Both Conditional-DDPM and the proposed PET-DPC generate images that are visually closer to QuanTOF, however, PET-DPC exhibits smaller residuals. Notably, both FastPET and Conditional-DDPM exhibit more pronounced quantitative errors in tumor regions. From the CNR analysis shown in Fig. 4, PET-DPC demonstrates tumor contrast most consistent with QuanTOF, while Conditional-DDPM performs moderately worse. In contrast, FastPET shows a clear reduction in tumor contrast compared to QuanTOF.

Table I summarizes the quantitative results across all 100 validation datasets, reported as mean $\pm$ standard deviation. Across all metrics, PET-DPC consistently outperforms both competing methods.

TABLE I
QUANTITATIVE COMPARISON OF PSNR, SSIM, AND NRMSE ACROSS 100
BRAIN VALIDATION DATASETS (MEAN ± STANDARD DEVIATION)

| Method | PSNR | SSIM | NRMSE |
|---|---|---|---|
| FastPET | 37.749 ± 2.111 | 0.941 ± 0.016 | 0.213 ± 0.103 |
| Conditional-DDPM | 44.782 ± 2.932 | 0.992 ± 0.003 | 0.098 ± 0.043 |
| PET-DPC | **47.703 ± 1.823** | **0.994 ± 0.002** | **0.066 ± 0.010** |

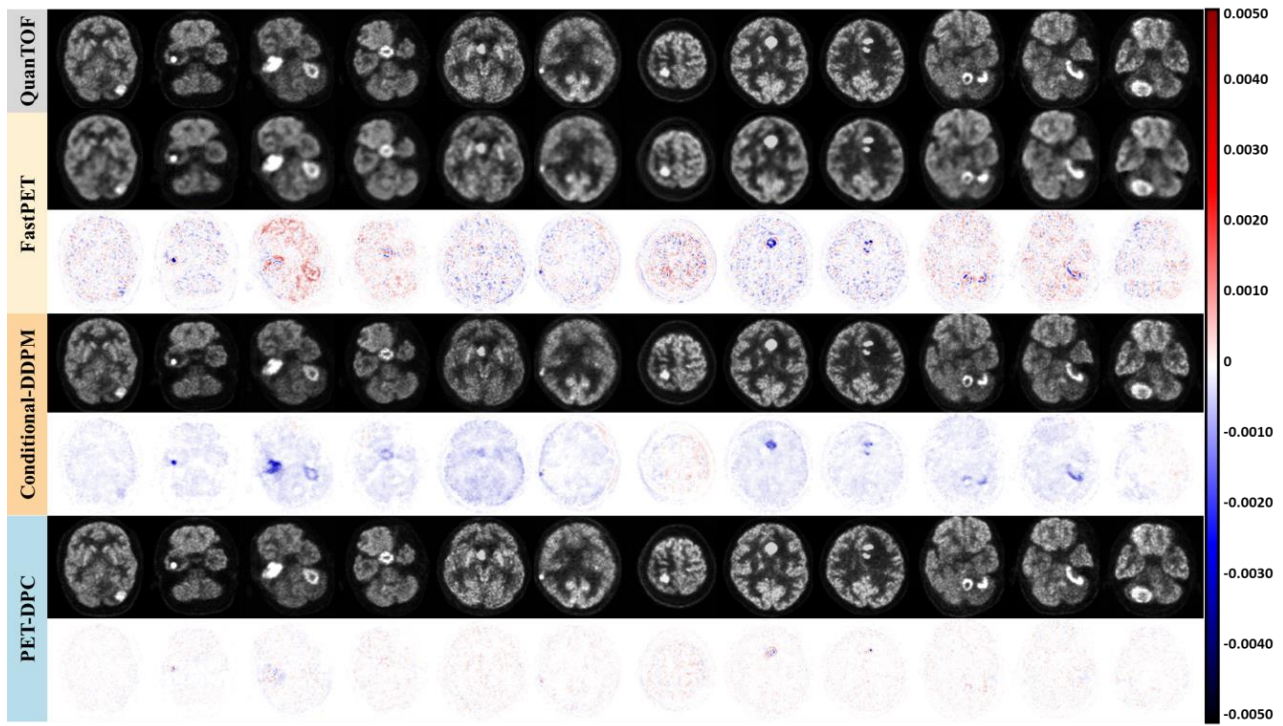

Fig. 2. Reconstructed brain images of 12 tumor cases using FastPET, Conditional-DDPM, and PET-DPC, along with corresponding residual maps relative to QuanTOF reconstructions.

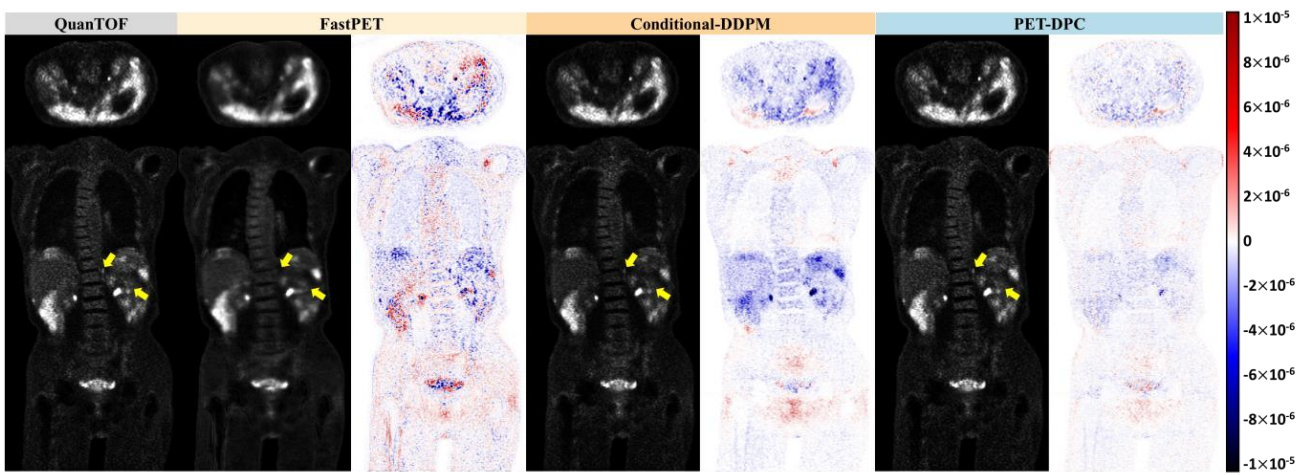

Fig. 3. Reconstructed whole-body images using FastPET, Conditional-DDPM, and PET-DPC, along with corresponding residual maps relative to QuanTOF reconstructions.

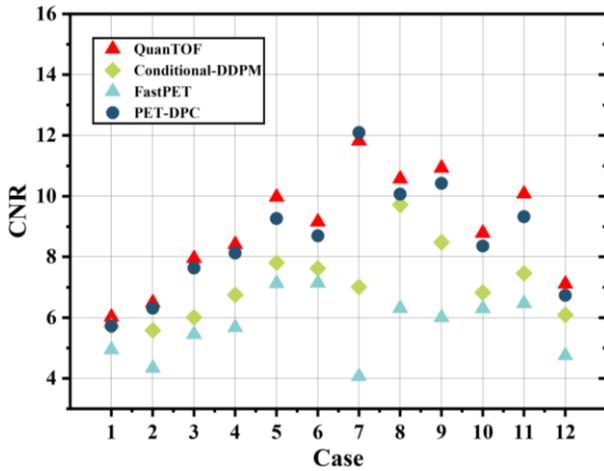

Fig. 4. Comparison of CNRs for reconstructed images obtained with FastPET, Conditional-DDPM, PET-DPC, and QuanTOF across 12 tumor cases.

### B. Clinical Whole-Body Study

A similar trend was observed in the five whole-body validation datasets. As shown in Fig. 3, PET-DPC consistently produced reconstructions with better structural preservation and reduced residuals, particularly in regions with high tracer uptake. In contrast, FastPET failed to recover some fine structures and small high-uptake regions, as indicated by the yellow arrows in Fig. 3. Quantitative results presented in Table II further confirm that PET-DPC outperforms both FastPET and Conditional-DDPM in terms of accuracy and robustness, even under the challenging conditions of whole-body imaging acquired with the CBM fast scanning protocol.

TABLE II
QUANTITATIVE COMPARISON OF PSNR, SSIM, AND NRMSE ON 5 WHOLE-BODY VALIDATION IMAGES (MEAN ± STANDARD DEVIATION)

| Method | PSNR | SSIM | NRMSE |
|---|---|---|---|
| FastPET | 38.012 ± 2.148 | 0.922 ± 0.012 | 0.663 ± 0.248 |
| Conditional-DDPM | 48.326 ± 3.455 | 0.988 ± 0.005 | 0.197 ± 0.032 |
| PET-DPC | **49.895 ± 2.873** | **0.991 ± 0.004** | **0.162 ± 0.022** |

### C. Physical phantom Study Results

To evaluate generalization on OOD data, the NEMA phantom was reconstructed using FastPET, Conditional-DDPM, and PET-DPC models trained on both brain and whole-body datasets. Fig. 5 shows the reconstructed images and the corresponding intensity profiles extracted from a representative line crossing the tumor region. End-to-end models without explicit physical correction (FastPET and Conditional-DDPM) suffer from noticeable background non-uniformity and reduced tumor uptake. In contrast, PET-DPC produces images with a more uniform background and a tumor-to-background activity ratio closer to the expected 4:1, closely matching the QuanTOF reference reconstruction.

Fig. 6 presents the CRC and BV results of the reconstructed NEMA phantom. Points closer to the top-left corner indicate better recovery of hot sphere uptake ratios and reduced noise. PET-DPC trained on whole-body datasets achieved higher CRC values than QuanTOF for all six hot spheres, with BV values either comparable or slightly higher. PET-DPC trained on brain datasets also produced results close to QuanTOF. In contrast, FastPET and Conditional-DDPM showed markedly higher BV and lower CRC, underscoring the substantial improvements provided by PET-DPC with posterior correction.

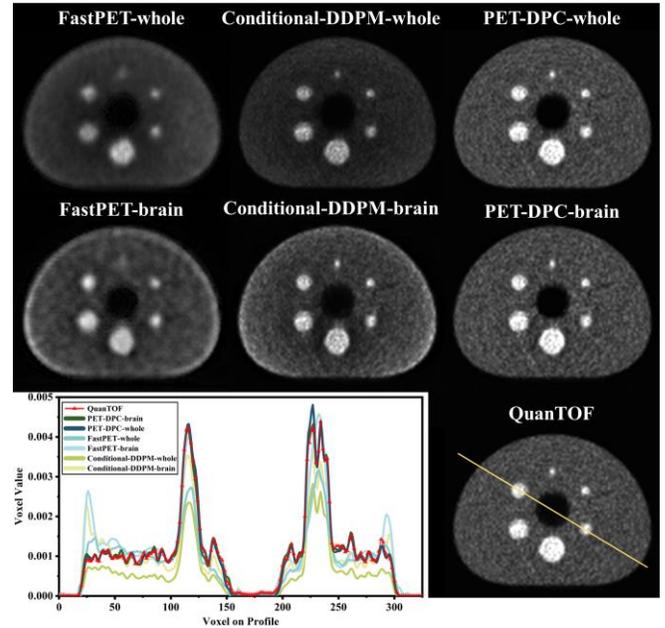

Fig. 5. Comparison of reconstructed NEMA phantom images obtained using FastPET, Conditional-DDPM, and PET-DPC (trained on brain and whole-body datasets), along with their corresponding intensity profiles extracted from a representative line passing through the tumor region.

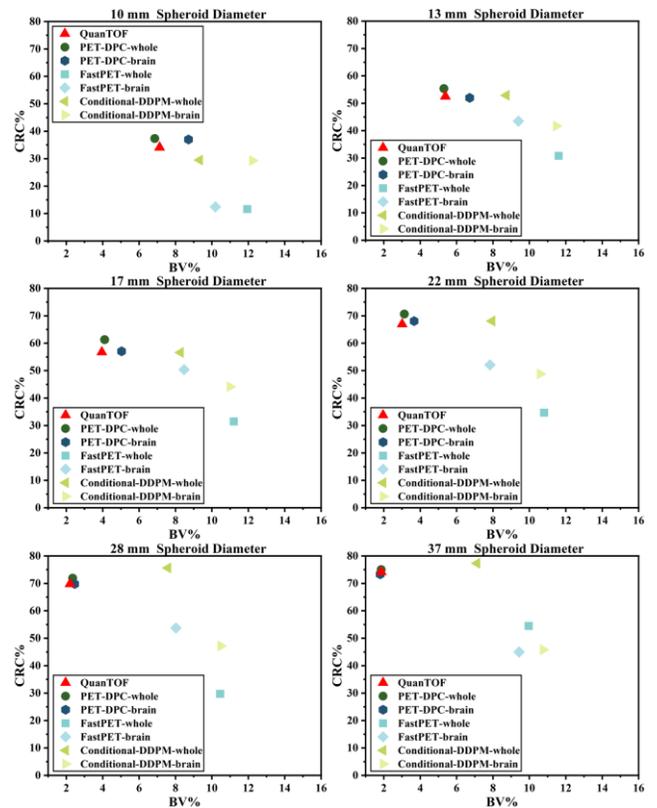

Fig. 6. Comparison of CRC and BV values for the six hot spheres across across seven reconstructed NEMA phantom images.

### D. Simulation Study

Fig. 7 shows simulated brain images reconstructed by FastPET, Conditional-DDPM, PET-DPC and QuanTOF, along with the reference activity image (ground truth). As the simulated data were generated with a predefined gray-to-white matter activity ratio, the mean pixel intensities of the corresponding regions were calculated across all 20 reconstructed images, and the resulting ratios are summarized in Table III.

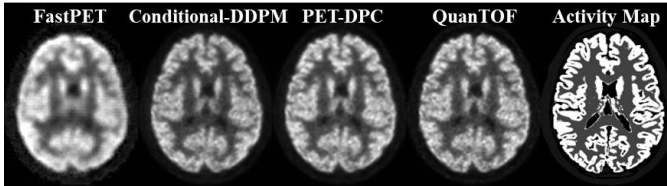

**Fig. 7.** Visual comparison of reconstructed simulated brain images using FastPET, Conditional-DDPM, PET-DPC and QuanTOF, along with the reference activity image.

The average gray-to-white matter ratios obtained from the QuanTOF and PET-DPC reconstructions were 1.898 and 1.868, respectively, compared to the reference ratio of 3.846, indicating strong quantitative consistency. To further evaluate noise suppression, CVs were calculated separately for gray and white matter regions. As shown in Table III, PET-DPC achieves effective noise reduction in both tissue types.

**TABLE III**
RESULTS OF GRAY-TO-WHITE MATTER ACTIVITY RATIOS AND CV ON SIMULATED BRAIN DATASETS

| Method | Gray / White Matter Ratio | Gray Matter CV（%） | White Matter CV（%） |
|---|---|---|---|
| Activity Map | **3.846 ± 0.000** | / | / |
| FastPET | 1.610 ± 0.034 | 20.637 ± 0.572 | 35.731 ± 1.169 |
| Conditional-DDPM | 1.853 ± 0.040 | 19.369 ± 0.487 | 34.320 ± 1.089 |
| QuanTOF | 1.898 ± 0.033 | 19.826 ± 0.474 | 32.930 ± 0.706 |
| PET-DPC | 1.868 ± 0.033 | **19.180 ± 0.473** | **32.648 ± 0.763** |

### E. Ablation Study

The advantage of incorporating physical information via posterior correction is evident in the comparison between PET-DPC and Conditional-DDPM. PET-DPC was implemented with five correction steps. To investigate the impact of the number of correction steps on the reconstruction of fine structures, this number was varied from 4 to 1 for brain datasets. Fig. 8 shows the resulting reconstructions and corresponding error maps relative to QuanTOF. As the number of correction steps increases, the discrepancy with QuanTOF decreases. Table IV presents quantitative results, confirming consistent improvement in reconstruction metrics with additional correction steps.

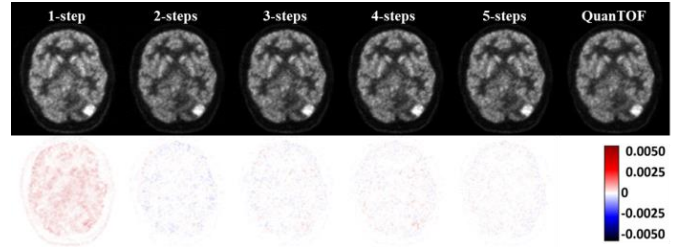

**Fig. 8.** Visual comparison of PET-DPC reconstructed images with varying correction steps, along with their corresponding residual maps relative to QuanTOF.

**TABLE IV**
COMPARISON OF PSNR, SSIM, AND NRMSE FOR 100 BRAIN VALIDATION IMAGES RECONSTRUCTED WITH DIFFERENT CORRECTION STEPS (MEAN ± STANDARD DEVIATION)

| Method | PSNR | SSIM | NRMSE |
|---|---|---|---|
| 5-steps | **47.703 ± 1.823** | **0.994 ± 0.002** | **0.066 ± 0.010** |
| 4-steps | 46.998 ± 1.749 | 0.994 ± 0.001 | 0.071 ± 0.009 |
| 3-steps | 46.354 ± 1.850 | 0.993 ± 0.002 | 0.077 ± 0.011 |
| 2-steps | 45.815 ± 2.482 | 0.992 ± 0.003 | 0.083 ± 0.020 |
| 1-step | 36.859 ± 2.220 | 0.966 ± 0.007 | 0.234 ± 0.022 |

### F. Computational cost

Table V summarizes the reconstruction times on brain and whole-body datasets, including the time required to generate the backprojected input images. Deep learning-based end-to-end methods (Conditional-DDPM and FastPET) significantly reduce reconstruction time by bypassing iterative computations in the list-mode data space, although this comes at the cost of quantitative accuracy. In contrast, PET-DPC incorporates physical information during sampling to preserve quantitative fidelity, which requires access to list-mode data and increases computation time. Nevertheless, PET-DPC still achieves significant time savings compared to QuanTOF. For instance, in a brain case with 58 million coincidence events, reconstruction time decreased from 82 s to 47 s, while in a whole-body case with 347 million events, it was reduced from 17,081 s (~4.7 hours) to 2,707 s (~45 minutes).

**TABLE V**
INFERENCE TIME (S) FOR DIFFERENT RECONSTRUCTION METHODS

| Algorithm | Brain data(58M) | Whole-body data(347M) |
|---|---|---|
| QuanTOF | 82 | 17081 |
| PET-DPC | 47 | 2707 |
| Conditional-DDPM | 30 | 242 |
| FastPET | 13 | 100 |

## V. DISCUSSION

In the brain reconstruction results, FastPET produces visually smoother images but fails to preserve fine structural details relative to the QuanTOF, resulting in pronounced error maps and substantially degraded quantitative metrics, including PSNR, SSIM, and NRMSE. In contrast, Conditional-DDPM, PET-DPC, and QuanTOF reconstructions are visually comparable, capturing consistent tumor morphology and structural details. Nevertheless, a noticeable difference in quantitative accuracy remains between Conditional-DDPM and PET-DPC.

For the whole-body datasets, FastPET similarly produces relatively smooth and low-noise images. Although this yields visually cleaner reconstructions in organs such as the liver and kidneys, FastPET either misses small high-uptake regions or produces structurally inconsistent results (highlighted by the yellow arrows), potentially compromising diagnostic reliability. Conditional-DDPM does not introduce visible artifacts but still underperforms PET-DPC in quantitative accuracy. These inaccuracies, particularly the failure to preserve subtle yet clinically relevant uptake regions, could result in misdiagnosis or missed diagnosis in clinical practice.

In deep learning–based PET reconstruction, supervised end-to-end methods such as FastPET aim to unify the entire reconstruction process, including all correction steps, into a single model. However, due to their limited interpretability, purely data-driven approaches may fail to accurately capture essential physical processes, such as scatter and attenuation, particularly when these are not explicitly modeled. Moreover, their strong dependence on training data often leads to poor generalization: reconstruction quality degrades when the input data distribution deviates from the training set. This limitation is evident in the NEMA phantom results. FastPET, Conditional-DDPM, and PET-DPC models were trained separately on brain and whole-body datasets and evaluated on NEMA phantom reconstructions. As shown in Fig. 5, FastPET and Conditional-DDPM exhibit pronounced background inhomogeneity, with darker central regions and abnormally bright edges—typical artifacts of reconstructions without proper attenuation and scatter correction. The 10 mm tumor is nearly indistinguishable in the FastPET result. Line profiles further highlight the substantial discrepancies between these models and the QuanTOF ground truth. In contrast, PET-DPC, whether trained on brain or whole-body datasets, reconstructs the phantom accurately after posterior correction, recovering the predefined 4:1 activity ratio. The comparison of CRC and BV values in Fig. 6 demonstrate that PET-DPC closely matches QuanTOF, with the PET-DPC model trained on whole-body datasets even surpassing QuanTOF in contrast recovery.

In the simulated dataset, the ground-truth gray-to-white matter activity ratio was 3.846:1. QuanTOF recovered a ratio of 1.898:1, while FastPET underestimated this value. Both Conditional-DDPM and PET-DPC produced ratios closer to QuanTOF, with PET-DPC yielding a slightly higher value. Additionally, the CV within gray and white matter regions indicates that PET-DPC achieves lower image noise than QuanTOF, highlighting its superior noise suppression.

In addition to quantitative accuracy, reconstruction time is a crucial metric in evaluating the clinical feasibility of PET reconstruction methods. Compared to QuanTOF, an iterative algorithm that is already optimized with GPU acceleration, our proposed PET-DPC significantly reduces reconstruction time while maintaining high quantitative accuracy. Unlike end-to-end deep learning methods that bypass physical modeling entirely, PET-DPC incorporates explicit physical corrections in each sampling step, which inevitably adds computational overhead and prolongs the overall reconstruction process. This trade-off ensures quantitative reliability, as rapid end-to-end models achieve faster reconstruction at the expense of reduced quantitative accuracy. Across both brain and whole-body datasets, the quantitative performance of FastPET and Conditional-DDPM is consistently inferior to that of PET-DPC. Moreover, these supervised learning models demonstrate limited generalizability to OOD data. For example, when applied to the NEMA phantom, both FastPET and Conditional-DDPM failed to accurately recover background uniformity and small high-uptake lesions, highlighting their limitations in robustness and clinical applicability.

As shown in Table V, for brain datasets, the inference time of PET-DPC is only marginally longer than that of Conditional-DDPM. In contrast, for whole-body datasets, PET-DPC requires more than ten times the inference time of Conditional-DDPM, mainly due to the complexity of scatter and attenuation corrections. Brain scans were acquired in a single bed position, whereas whole-body data were collected in CBM mode. Scatter and attenuation correction in CBM mode is more challenging because the attenuation map (μ-map) changes dynamically. During reconstruction, every 10 s of acquisition was treated as a single bed position for scatter and attenuation correction. Using longer scan segments per correction step could further improve PET-DPC reconstruction speed.

In our implementation, PET-DPC uses five sampling steps, each followed by a physics-based correction. As shown in the ablation studies, increasing the number of sampling and correction steps leads to further improvements in quantitative accuracy but at the cost of longer reconstruction time. This highlights a fundamental trade-off between reconstruction quality and computational efficiency, which must be carefully balanced for deployment in clinical practice.

By explicitly embedding physics-based corrections within the end-to-end reconstruction pipeline, PET-DPC provides a viable solution for achieving both accelerated reconstruction and high quantitative accuracy. Nevertheless, reconstruction time remains a key area for future improvement. In our current implementation, the primary computational bottleneck lies in the physics-based correction performed in data space after each sampling step. Although the input to each correction stage is already a close approximation of the final reconstruction, this step still requires loading the full list-mode coincidence data and applying attenuation and scatter corrections. Despite GPU acceleration, this correction process remains time-consuming. Moving forward, we plan to explore more efficient correction strategies to further reduce overall reconstruction time and improve the clinical feasibility of the approach.

## VI. CONCLUSION

In this work, we proposed a conditional diffusion model with posterior physical correction, PET-DPC, to address the quantitative accuracy limitations of end-to-end PET reconstruction methods that omit explicit modeling of physical processes. PET-DPC effectively bridges the gap between data-driven speed and physics-based accuracy in PET reconstruction. By embedding physical corrections within a conditional diffusion framework via posterior sampling, it achieves quantitatively reliable images comparable to gold-standard iterative methods while significantly accelerating processing. The model's robustness is validated across diverse clinical, phantom and simulated datasets, demonstrating superior generalization over purely end-to-end approaches. Key innovations include GTP-image generation and iterative posterior correction, which mitigate artifacts and intensity mismatches. Though computational overhead remains a challenge (especially for whole-body CBM scans), PET-DPC offers a clinically viable balance between reconstruction quality and speed. Future work will optimize correction efficiency and expand clinical validation.